%
%
\def\today{\ifcase\month\or January\or February\or March\or April\or May\or
June\or July\or August\or September\or October\or November\or December\fi
\space\number\day, \number\year}
%
%
\newcount\notenumber

\def\note{\global\advance\notenumber by 1 \footnote{$^{\the\notenumber}$}}
%
%
\newif\ifsectionnumbering
\newcount\eqnumber
\def\cleareqnumber{\eqnumber=0}
\def\numbereq{\global\advance\eqnumber by 1
\ifsectionnumbering\eqno(\the\secnumber.\the\eqnumber)\else\eqno
(\the\eqnumber)\fi}
\def\eqalinno{{\global\advance\eqnumber by 1}
\ifsectionnumbering(\the\secnumber.\the\eqnumber)\else(\the\eqnumber)\fi}
\def\name#1{\ifsectionnumbering\xdef#1{\the\secnumber.\the\eqnumber}
\else\xdef#1{\the\eqnumber}\fi}
\def\nosectionnumbering{\sectionnumberingfalse}
\sectionnumberingtrue
%
%
\newcount\refnumber

\immediate\openout1=refs.tex
\immediate\write1{\noexpand\frenchspacing}
\immediate\write1{\parskip=0pt}
\def\ref#1#2{\global\advance\refnumber by 1%
[\the\refnumber]\xdef#1{\the\refnumber}%
\immediate\write1{\noexpand\item{[#1]}#2}}
\def\tie{\noexpand~}

%
%
\font\twelvebf=cmbx10 scaled \magstep1
\newcount\secnumber

\def\newsection#1.{\ifsectionnumbering\cleareqnumber\else\fi%
	\global\advance\secnumber by 1%
	\bigbreak\bigskip\par%
	\line{\twelvebf \the\secnumber. #1.\hfil}\nobreak\medskip\par\noindent}
%
%
%
\def \sqr#1#2{{\vcenter{\vbox{\hrule height.#2pt
	\hbox{\vrule width.#2pt height#1pt \kern#1pt
		\vrule width.#2pt}
		\hrule height.#2pt}}}}

%
%
%
\newdimen\fullhsize
\def\fiddle{\fullhsize=6.5truein \hsize=3.2truein}
\def\fullline{\hbox to\fullhsize}
\def\mkhdline{\vbox to 0pt{\vskip-22.5pt
	\fullline{\vbox to8.5pt{}\the\headline}\vss}\nointerlineskip}
\def\mkftline{\baselineskip=24pt\fullline{\the\footline}}
\let\lr=L \newbox\leftcolumn
\def\twocolumns{\fiddle
	\output={\if L\lr \global\setbox\leftcolumn=\columnbox
		\global\let\lr=R \else \doubleformat \global\let\lr=L\fi
		\ifnum\outputpenalty>-20000 \else\dosupereject\fi}}
\def\doubleformat{\shipout\vbox{\mkhdline
		\fullline{\box\leftcolumn\hfil\columnbox}
		\mkftline} \advancepageno}
\def\columnbox{\leftline{\pagebody}}
\nosectionnumbering
\magnification=1200
\def\pr#1 {Phys. Rev. {\bf D#1\tie }}
\def\pe#1 {Phys. Rev. {\bf #1\tie}}
\def\pre#1 {Phys. Rep. {\bf #1\tie}}
\def\pl#1 {Phys. Lett. {\bf #1B\tie }}
\def\prl#1 {Phys. Rev. Lett. {\bf #1\tie }}
\def\np#1 {Nucl. Phys. {\bf B#1\tie }}
\def\ap#1 {Ann. Phys. (NY) {\bf #1\tie }}
\def\cmp#1 {Commun. Math. Phys. {\bf #1\tie }}
\def\imp#1 {Int. Jour. Mod. Phys. {\bf A#1\tie }}
\def\mpl#1 {Mod. Phys. Lett. {\bf A#1\tie}}
\def\jhep#1 {JHEP {\bf #1\tie}}
\def\nuo#1 {Nuovo Cimento {\bf B#1\tie}}
\def\tie{\noexpand~}

\parskip=15pt plus 4pt minus 3pt
\headline{\ifnum \pageno>1\it\hfil Linearized Gravity in
Isotropic Coordinates $\ldots$\else \hfil\fi}
\font\title=cmbx10 scaled\magstep1
\font\tit=cmti10 scaled\magstep1
\footline{\ifnum \pageno>1 \hfil \folio \hfil \else
\hfil\fi}
\raggedbottom


\overfullrule0pt


\rightline{\vbox{\hbox{RU01-05-B}\hbox{hep-th/0103265}}}
\vfill
\centerline{\title LINEARIZED GRAVITY IN ISOTROPIC COORDINATES}
\centerline{\title IN THE BRANE WORLD}
\vfill
{\centerline{\title Ioannis Giannakis
and Hai-cang Ren \footnote{$^{\dag}$}
{\rm e-mail: \vtop{\baselineskip12pt
\hbox{giannak@theory.rockefeller.edu, ren@theory.rockefeller.edu,}}}}
}
\medskip
\centerline{{\tit Physics Department, The Rockefeller
University}}
\centerline{\tit 1230 York Avenue, New York, NY
10021-6399}
\vfill
\centerline{\title Abstract}
\bigskip
{\narrower\narrower
We solve the Einstein equations in the Randall-Sundrum
framework using an isotropic {\it ans\"atz\/} for
the metric and obtain
an exact expression to first order in the gravitational
coupling. The solution is free from metric singularities
away from the source and it satisfies the Israel matching
condition on a straight brane. At distances far away from the
source and on the {\it physical brane} this solution coincides
with the 4-D Schwarzschild metric in isotropic coordinates.
Furthermore we show that the extension of the standard
Schwarzschild horizon in the bulk is tubular for any diagonal
form of the metric while there is no restriction for
the extension of the Schwarzschild horizon in isotropic coordinates.
\par}
\vfill\vfill\break


\newsection Introduction.

Recently there have been several attempts to achieve
localization of gravity.
In the brane world ordinary matter and its gauge
interactions are confined within a 4-D hypersurface,
referred to subsequently as the {\it physical brane}.
The graviton, however, is allowed to propagate in extra space
dimensions.
One implementation of such a brane world scenario was
proposed by Randall and Sundrum \ref{\randall}{
L. Randall and R. Sundrum, \prl83 (1999) 4690.}
within the framework of General Relativity.
The {\it physical brane} in their model is the junction
of two pieces of 5-D spacetime manifolds that
are asymptotically {\it anti-de Sitter}. The Gauss-normal form
of the metric in this space is
$$
ds^2=e^{-2\kappa|y|}{\bar g}_{\mu\nu}dx^\mu dx^\nu+dy^2,
\numbereq\name{\eqena}
$$
with the brane located at $y=0$ and $\kappa > 0$ sets
the energy scale of the extra space dimension. The metric
${\bar g}_{\mu\nu}$ is determined by the 5-D Einstein equations
$$
R_{mn}-{1\over 2}Rg_{mn}-\Lambda g_{mn}=-4{\pi}^2G_5
T_{\mu\nu}\delta^{\mu}_{m}\delta^{\nu}_{n}{\delta}(y)
+6{\kappa}g_{\mu\nu}\delta^{\mu}_{m}\delta^{\nu}_{n}{\delta}(y)
\numbereq\name{\eqdyo}
$$
where the cosmological constant $\Lambda=-6\kappa^2$,
$G_5$ is the 5-D gravitational constant and $T_{\mu\nu}$
the energy-momentum tensor on the brane. Here and
throughout the 
paper, we adopt the convention that the Greek indices take values 
0-3 and the Latin indices 0-4. The 4-D gravitational
constant is given by $G \sim {{G_5}\over {\kappa}}$.
In the absence of matter, $T_{\mu\nu}=0$, ${\bar g}_{\mu\nu}=
\eta_{\mu\nu}$ is a solution of equation (\eqdyo).
Subsequently, the metric in equation (\eqena) becomes
that of $AdS_5$. The solution to equation (\eqdyo) can
also be obtained from the solution to the sourceless equation,
$T_{\mu\nu}=0$, subject to the appropriate Israel matching condition
\ref{\isr}{W. Israel, \nuo44
(1966) 1.} determined by $T_{\mu\nu}$.
The perturbative solution of (\eqdyo) to the linear
order in $G$ for an arbitrary $T_{\mu\nu}$ 
\ref{\giddings}{S. Giddings, E. Katz
and L. Randall, \jhep0003 (2000) 023.}
and to second
order $G^2$ for a static spherical mass distribution on
the brane \ref{\ren}{I. Giannakis and H. C. Ren, \pr63 (2001)
024001.}
reveals no tangible disagreement with the
classical tests of 4-D General Relativity at large distances, i.e.,
$\kappa r>>1$. Discussions on different aspects of
linearized gravity in the Randall-Sundrum framework appeared
in \ref{\lin}{I. Y. Arefe'va, M. Ivanov, W. Muck, K. S. Viswanathan,
and I. V. Volovich, \np590 (2000) 273; M. Ivanov and I. V. Volovich,
{\it Metric fluctuations in brane worlds}, hep-th/9912242; H. Collins
and B. Holdom, \pr62 (2000) 124008;
W. Muck, K. S. Viswanathan
and I. V. Volovich, \pr62 (2000) 105019;
R. Dick and D. Mikulovicz, \pl476 (2000) 363.}.

A drawback of the Gauss-normal coordinates is the failure
to compromise between a straight brane (i.e. Israel
matching condition at $y=0$) and a non-singular boundary
condition for the metric ${\bar g}_{\mu\nu}$ as $y \to \infty$.
This was shown explicitly in the weak field approximation
( to first order [\randall] or to quadratic order [\ren]
in the gravitational coupling )
for which the deviation of ${\bar g}_{\mu\nu}$ from its
Minkowski value $\eta_{\mu\nu}$, i.e. $h_{\mu\nu}=
{\bar g}_{\mu\nu}-\eta_{\mu\nu}$, grows exponentially with
increasing $y$ if the Israel matching condition is imposed
at $y=0$. On the other hand, if the condition
$\lim_{y\to \infty}h_{\mu\nu}=0$ is imposed,
the Israel matching condition fails to hold at $y=0$
and the brane is bent. Beyond the weak field approximation,
the lack of compromise is reflected in the rigorous statement
that the $5-D$ extension of the Schwarzschild horizon of
a physical black hole is always tubular, parallel to
the $y$-axis.

It was suggested in \ref{\kaku}{Z. Kakushadze, \pl497 (2001) 125.}
that the Israel matching condition at $y=0$ and
a non-singular boundary condition for the metric as
$y \to \infty$ can be achieved  by introducing
the graviscalar component of the metric ( the coefficient
in front of $dy^2$ in the metric
{\it ans\"atz\/}). In the present paper we explore
this possibility for a physical black hole.
By using a particular diagonal metric {\it ans\"atz\/}
we find an explicit solution to the linearized Einstein equations
which satisfies the Israel matching condition (straight brane)
while behaves well at the $AdS_5$ horizon ($y \to \infty$).
Since the metric on the brane approximates the $4-D$
Schwarzschild metric in isotropic coordinates for
$\kappa {\rho}>>1$ and $\kappa GM >>1$
$$
ds^2 \cong - \Big ( {{1-{GM\over {2\rho}}}\over {1+{GM\over {2\rho}}}}
\Big )^2dt^2+ \Big( 1+{GM\over {2\rho}} \Big)^4
\Big(d{\rho}^2+{\rho}^2d{\Omega}^2 \Big),
\numbereq\name{\eqxaristeas}
$$ 
we shall refer
to these $5-D$ coordinates as the isotropic coordinates
( this is not to be confused with any Gauss-normal
coordinates which approximate the $4-D$ isotropic
Schwarzschild metric). Beyond the linear approximation
we demonstrate that the $5-D$ extension of the horizon
in the isotropic coordinates is closed and might be of
a pancake shape. On the other hand if one requires the
standard Schwarzshild metric to be implemented on the
{\it physical brane}
$$
ds^2 \cong - \Big ( {1-{2GM\over r}}
\Big )dt^2+ \Big( 1-{2GM\over r} \Big)^{-1}
dr^2+r^2d{\Omega}^2,
\numbereq\name{\eqliolidhs}
$$ 
a straight brane and a non-singular
boundary condition at the $AdS_5$ horizon can never be
compromised with any diagonal metric {\it ans\"atz\/},
since the off brane extension of the horizon in this
case is always tubular.

In the next section, we shall present the explicit
solution to the linearized Einstein equations in the
isotropic coordinates. The rigorous statements on the
horizon with an arbitrary diagonal metric {\it ans\"atz\/} will
be discussed in section 3. In the final section,
we shall summarize our results and speculate on the
form of the solution in isotropic coordinates beyond
the linear approximation. For the benefit of the readers,
we present in the appendix the explicit form of all non-zero
components of the Ricci tensor with a diagonal metric {\it ans\"atz\/}.

\newsection Linearized Solution in Isotropic Coordinates.

The  5-D Einstein sourceless equations can be rewritten as
$$
R_{\mu\nu}-4\kappa^2g_{\mu\nu}=0, \quad R_{y\mu}=0,
\quad R_{yy}-4\kappa^2=0.
\numbereq\name{\eqtesse}
$$
The most general metric in $D=4+1$ dimensions produced by a static,
spherically symmetric matter distribution on the {\it physical brane}
or equivalently, axially symmetric in the bulk can always be 
brought to Gauss normal form:
$$
ds^2=e^{-2\kappa|y|}(-e^adt^2+e^bdr^2+e^cr^2d\Omega^2)+dy^2,
\numbereq\name{\eqpente}
$$
where $d\Omega^2=d\theta^2+\sin^2\theta d\phi^2$ is the solid angle on 
$S^2$ and $a$, $b$ and $c$ are functions of $r$ and $y$.

In this paper we will consider an alternative form of the metric
in $D=4+1$, written in isotropic coordinates
$$
ds^2=e^{-2\kappa\eta}(-e^{\alpha}dt^2
+e^{\beta}(d{\rho}^2+{\rho}^2d\Omega^2))+e^fd{\eta}^2,
\numbereq\name{\eqpene}
$$
where $\alpha$, $\beta$ and $f$ are functions of $\rho$ and $\eta$,
and we will obtain an exact solution of the Einstein equations
to first order in the gravitational coupling G.
Since the linearized solution is known in the Gauss normal
coordinates [\ren]
our objective initially would be to determine the form of
the coordinate transformations that transform the Gauss normal
form of the metric, equation (\eqpente), to the
isotropic one, equation (\eqpene).

Let's perform a coordinate transformation (gauge transformation) generated
by $u, v$, functions of $\rho$ and $\eta$ such that
$r=\rho+u(\rho, \eta)$ and $y=\eta+v(\rho, \eta)$.
We find that the Gauss-normal form of the metric transforms 
to linear order as follows
$$
\eqalign{
ds^2=e^{-2\kappa{\eta}}[-(1+a-2{\kappa}v)dt^2&+
(1+b+2u^{\prime}-2{\kappa}v)d{\rho}^2+(1+c-2{\kappa}v
+{2u\over {\rho}}){\rho}^2d\Omega^2]\cr
&+(1+2{\dot v})d{\eta}^2
+2(e^{-2\kappa{\eta}}{\dot u}+v^{\prime})d{\rho}d{\eta}, \cr}
\numbereq\name{\eqtevek}
$$
where $u^\prime$ and $\dot u$ indicate differentiation
with respect to $\rho$ and $\eta$ respectively.
Since we would like to transform the metric from the Gauss-normal
form to the isotropic form we seek appropriate $u$ and $v$
such that they satisfy
$$
e^{-2\kappa{\eta}}\dot u+v^\prime=0, \qquad b+2u^\prime=
c+{2u\over\rho}.
\numbereq\name{\eqtevekidhs}
$$
Furthemore with the identification
$$
\alpha=a-2{\kappa}v, \qquad \beta=b-2{\kappa}v+2u^\prime,
\qquad f=2\dot v
\numbereq\name{\eqvekid}
$$
we transform the metric into the isotropic form
$$
ds^2=e^{-2\kappa{\eta}}[-(1+\alpha)dt^2+
(1+\beta)(d{\rho}^2+{\rho}^2d\Omega^2)]+(1+f)d{\eta}^2.
\numbereq\name{\eqexi}
$$

Substituting the metric (\eqexi) into
equations (\eqtesse), we obtain the following
components of the linearized Einstein equation outside the source:
$$
R_{tt}+4\kappa^2 e^{-2\kappa{\eta}+\alpha}=
-{1\over 2}{\alpha}^{\prime\prime}
-{1\over{\rho}}{\alpha}^\prime-{1\over 2}
e^{-2\kappa{\eta}}\Big[\ddot {\alpha}-
5\kappa\dot {\alpha}-3\kappa\dot {\beta}+\kappa
\dot f-8{\kappa}^2f \Big]=0
\numbereq\name{\eqevon}
$$
$$
R_{\rho\rho}-4\kappa^2e^{-2\kappa{\eta}+\beta}={1\over 2}
{\alpha}^{\prime\prime}+{1\over 2}f^{\prime\prime}
+{\beta}^{\prime\prime}
+{1\over{\rho}}{\beta}^\prime
+{1\over 2}e^{-2\kappa{\eta}}\Big[\ddot{\beta}-7\kappa\dot{\beta}
-\kappa\dot{\alpha}+\kappa\dot f-8{\kappa}^2f \Big]=0
\numbereq\name{\eqasho}
$$
$$
R_{\theta\theta}-4\kappa^2 {\rho}^2e^{-2\kappa{\eta}+\beta}
={1\over 2}{\rho}^2\Big[
{\beta}^{\prime\prime}+{3\over {\rho}}{\beta}^\prime
+{1\over {\rho}}{\alpha}^\prime+{1\over {\rho}}f^\prime
+e^{-2\kappa{\eta}}[
\ddot{\beta}-\kappa(\dot{\alpha}-\dot f)-7\kappa \dot{\beta}
-8{\kappa}^2f] \Big]=0
\numbereq\name{\eqhbcv}
$$
$$
\eqalign{
R_{\eta\eta}&-4\kappa^2e^f=
{1\over 2}e^{2{\kappa}{\eta}}(f^{\prime\prime}
+{2\over {\rho}}f^{\prime})+{1\over 2}(\ddot{\alpha}+3\ddot{\beta})
-\kappa\dot{\alpha}-3{\kappa}\dot{\beta}+2{\kappa}\dot f+8{\kappa}^2f=0\cr
&R_{\rho\eta}=R_{\eta\rho}={1\over 2}\dot {\alpha}^\prime
+\dot {\beta}^{\prime}+{3\over 2}{\kappa}f^{\prime}=0,\cr}
\numbereq\name{\eqefta}
$$
These equations apply 
to the positive side of the brane, ${\eta}>0$, the corresponding
equations to the negative side of the brane, ${\eta}<0$, are obtained
by switching the sign of $\kappa$.
 
In reference [\ren] solutions to the linearized Einstein
equations using the Gauss-normal form of the metric, equation (\eqena),
were obtained in two different coordinate
systems, or equivalenty, in two different
gauges. The solution in the coordinate
system based on the $AdS$ horizon is free from
metric singularities far away from the source but
fails to satisfy the Israel matching condition at $y=0$, while
when it is transformed to a coordinate
system based on the brane it satisfies the
Israel matching condition at $y=0$ but is not
free from metric singularities anymore. In this
coordinate system the {\it physical brane} appears bent
\ref{\gar}{J. Garriga and
T. Tanaka, \prl84 (2000) 2778.}.
In the latter case the solution is given by the expression
$$
\eqalign{
a^{P}(r, \zeta)&=-{{8GM{\kappa}}\over {3{\pi}}}
{\zeta}^2\int_0^\infty dppj_0(pr)
{{{K_2(p{\zeta})}}\over {K_1(\hat p)}}+{2GM\over {3r}}\cr
b^{P}(r, \zeta)&={{8GM{\kappa}}\over {3{\pi}}}{\zeta}^2\int_0^\infty dpp
{{j_1(pr)}\over {pr}}
{{{K_2(p{\zeta})}}\over {K_1(\hat p)}}+{2GM\over {3r}}+{{2GM}\over
{3r^3}}{\zeta}^2\cr
c^{P}(r, \zeta)&={{8GM{\kappa}}\over {3{\pi}}}{\zeta}^2\int_0^\infty dpp
{1\over 2}[j_0(pr)-{{j_1(pr)}\over {pr}}]
{{K_2(p{\zeta})}\over {K_1(\hat p)}}-{2GM\over {3r}}-{{GM}\over
{3r^3}}{\zeta}^2\cr}
\numbereq\name{\eqrvzu}
$$
where we have introduced $\hat p={p\over {\kappa}}$,
$\zeta={1\over {\kappa}}e^{{\kappa}y}$,
$K_\nu(\zeta)$ is the modified 
Bessel function of the second kind and $j_0(x)$ is the spherical Bessel 
function.
The superscript P indicates that $a^{P}$, $b^{P}$ and
$c^{P}$ satisfy the Neumann boundary condition
( Israel matching condition away from the source )
on the brane located at $\zeta={1\over {\kappa}}$
$$
{{\partial}\over {\partial {\zeta}}}a^{P}|_{\zeta={1\over {\kappa}}}=
{{\partial}\over {\partial {\zeta}}}b^{P}|_{\zeta={1\over {\kappa}}}=
{{\partial}\over {\partial {\zeta}}}c^{P}|_{\zeta={1\over {\kappa}}}=0.
\numbereq\name{\eqrviob}
$$
We would like to find an exact solution to the linearized
Einstein equations,
expressed in the isotropic coordinates, equations (\eqevon)-(\eqefta).
We shall proceed by determing the form of the
parameters of coordinate transformations $u$ and $v$
that transform the metric from the Gauss-normal form to the
isotropic one. The second of equations (\eqtevekidhs) can be solved
and provides us with the following expression for $u$
$$
u(\rho, \zeta)=-{\rho}\int_{\rho}^{\infty}{{b^{P}-c^{P}}\over {2s}}ds
+{\rho}{\chi}(\zeta)
\numbereq\name{\eqbios}
$$
where $\chi$ is an arbitrary function of $\zeta$.
Using the expressions for $b^{P}$ and $c^{P}$ from equations
(\eqrvzu) we find the following expression for $u$ and
consequently for $v$ by substituting into the first of
equations (\eqtevekidhs)
$$
\eqalign{
u(\rho, \zeta)&={\rho}{\chi}(\zeta)+{{2GM}\over 3}
+{{GM{\zeta}^2}\over {6{\rho}^2}}
+{{2GM{\kappa}}\over {3{\pi}}}{\zeta}^2\int_0^\infty dp
{j_1(p{\rho})}{{{K_2(p{\zeta})}}\over {K_1(\hat p)}}\cr
v(\rho, \zeta)&={\phi}(\zeta)+{{GM}\over {3{\kappa}{\rho}}}-
{{{\rho}^2}\over {2{\kappa}{\zeta}}}{{d\chi}\over {d\zeta}}
-{{2GM}\over {3{\pi}}}{\zeta}\int_0^\infty dp
{j_0(p{\rho})}{{{K_1(p{\zeta})}}\over {K_1(\hat p)}}\cr}
\numbereq\name{\eqsphere}
$$
where $\phi$ is another arbitrary function of $\zeta$.

Consequently we substitute equation (\eqsphere) into
equation (\eqvekid)
and demand that the expressions for $\alpha$, $\beta$ and $f$
are free from metric singularities far away from the source
$$
\lim_{\zeta, \rho\to\infty}\alpha=\lim_{\zeta, \rho\to\infty}\beta
=\lim_{\zeta, \rho\to\infty}f=0.
\numbereq\name{\eqexy}
$$
We find that $\chi(\zeta)=\phi(\zeta)=0$ and consequently
the solution takes the form
$$
\eqalign{
\alpha(\rho, \zeta)&=-{{8GM{\kappa}}\over {3{\pi}}}
{\zeta}^2\int_0^\infty dpp
{j_0(p{\rho})}{{{K_2(p{\zeta})}}\over {K_1(\hat p)}}+
{{4GM{\kappa}}\over {3{\pi}}}
{\zeta}\int_0^\infty dp
{j_0(p{\rho})}{{{K_1(p{\zeta})}}\over {K_1(\hat p)}}\cr
\beta(\rho, \zeta)&={{4GM{\kappa}}\over {3{\pi}}}
{\zeta}^2\int_0^\infty dpp
{j_0(p{\rho})}{{{K_2(p{\zeta})}}\over {K_1(\hat p)}}+
{{4GM{\kappa}}\over {3{\pi}}}
{\zeta}\int_0^\infty dp
{j_0(p{\rho})}{{{K_1(p{\zeta})}}\over {K_1(\hat p)}}\cr
f(\rho, \zeta)&={{4GM{\kappa}}\over {3{\pi}}}
{\zeta}^2\int_0^\infty dpp
{j_0(p{\rho})}{{{K_0(p{\zeta})}}\over {K_1(\hat p)}}\cr}.
\numbereq\name{\eqartis}
$$
In this coordinate system the brane remains straight
since $v(\rho,{1\over {\kappa}})=0$, in contrast
with the Gauss-normal coordinates in which the
brane was bent in the system in which the solution
was free of singularities off brane.

We still need to check whether the solution
$\alpha(\rho, \zeta)$, $\beta(\rho, \zeta)$ and
$f(\rho, \zeta)$ satisfies the equations of motion
(\eqevon)-(\eqefta).
Lets define
$$
\eqalign{
\Phi(\rho, \zeta)&={{4GM{\kappa}}\over {3{\pi}}}
{\zeta}^2\int_0^\infty dpp
{j_0(p{\rho})}{{{K_2(p{\zeta})}}\over {K_1(\hat p)}}\cr
\Psi(\rho, \zeta)&={{4GM{\kappa}}\over {3{\pi}}}
{\zeta}\int_0^\infty dp
{j_0(p{\rho})}{{{K_1(p{\zeta})}}\over {K_1(\hat p)}}\cr}
\numbereq\name{\eqartistic}
$$
such that
$$
\alpha(\rho, \zeta)=-2\Phi(\rho, \zeta)+\Psi(\rho, \zeta),
\quad \beta(\rho, \zeta)=\Phi(\rho, \zeta)+\Psi(\rho, \zeta),
\quad f(\rho, \zeta)=-{\zeta}{{\partial\Psi}\over {\partial\zeta}}
(\rho, \zeta).
\numbereq\name{\eqerion}
$$
Equation (\eqevon) then becomes
$$
-2\Phi^{\prime\prime}+\Psi^{\prime\prime}+{2\over {\rho}}(-2
\Phi^{\prime}+\Psi^{\prime})-2
{{{\partial^2}\Phi}\over {\partial{\zeta^2}}}
+{5\over {\zeta}}{{\partial\Phi}\over {\partial\zeta}}=0
\numbereq\name{\eqarite}
$$
Subsequently we substitute the expressions
for the solution into equation (\eqarite) and
taking into account that
$x^2K_{2}^{\prime\prime}(x)+xK_{2}^{\prime}(x)=
(4+x^2)K_{2}(x)$ together with $[x^2K_2(x)]^\prime=-x^2K_1(x)$, we verify
that they satisfy equation (\eqevon). Similarly we demonstrate
that the expressions for $\alpha$, $\beta$ and $f$ satisfy
the remaining Einstein equations.

If either $\rho$ or $\zeta\equiv{1\over\kappa}e^{\kappa y}$
becomes large, i.e, 
$\kappa {\rho}>>1$ or $\kappa {\zeta}>>1$, the integrals (\eqartistic),
are dominated by the region
where $\hat p<<1$. The modified Bessel function in the
denominator, $K_1(\hat p)\simeq {1\over\hat p}$, and the
integrals can be carried out explicitly. We find that
$$
\Phi(\rho, \zeta)={{2GM}\over 3}{{2{\rho}^2+3{\zeta}^2}
\over {({\rho}^2
+{\zeta}^2)^{3\over 2}}} \qquad
\Psi({\rho}, \zeta)={{2GM}\over 3}{1\over\sqrt{{\rho}^2+\zeta^2}}
\numbereq\name{\eqaoibcv}
$$
which leads to the following approximate expressions for
$\alpha$, $\beta$
and $f$
$$
\alpha(\rho, \zeta)=-{{2GM}\over 3}{{3{\rho}^2+5{\zeta}^2}
\over {({\rho}^2
+{\zeta}^2)^{3\over 2}}}, \quad
\beta(\rho, \zeta)={{2GM}\over 3}{{3{\rho}^2+4{\zeta}^2}
\over {({\rho}^2
+{\zeta}^2)^{3\over 2}}}, \quad
f(\rho, \zeta)={{4GM}\over 3}{{{\zeta}^2}
\over {({\rho}^2
+{\zeta}^2)^{3\over 2}}}.
\numbereq\name{\eqskend}
$$
It is straightforward to verify that the
metric on the {\it physical plane}
$(\zeta={1\over {\kappa}})$ becomes
for  $\rho>>1/\kappa$
$$
ds^2=-(1-{2GM\over {\rho}}+\cdots)dt^2+
(1+{{2GM}\over {\rho}}+\cdots)(d{\rho}^2+{\rho}^2d\Omega^2),
\numbereq\name{\eqtud}
$$
thus reproducing the standard form of the Schwarzschild metric
in isotropic coordinates.
The dots in equation (\eqtud) represent terms of order $O({{G^2M^2}
\over {\rho}^2})$
and higher.
The weak field expansion of a general static spherical metric in its
isotropic form which is not
necessarily determined by the Einstein equations
is \ref{\wei}{S. Weinberg, {\it Gravitation and Cosmology,
Chapter 8}, John Wiley and Sons, Inc. (1972).}
$$
ds^2=-(1-2{\alpha}_1{GM\over {\rho}}
+2\alpha_2{G^2M^2\over {\rho}^2}+...)dt^2
+(1+2{\alpha}_3{GM\over {\rho}}+...)(d{\rho}^2+{\rho}^2d\Omega^2).
\numbereq\name{\eqfgih}
$$
General Relativity predicts that $\alpha_1=\alpha_2=\alpha_3=1$.
Comparing equations (\eqtud) and (\eqfgih) we note that the
Randall-Sundrum scenario is consistent with all the
experimental tests of linearized General Relativity.

Let's now check whether our solution $\alpha$, $\beta$
and $f$ satisfies the Israel matching condition on the
brane.
The Israel matching condition in the case of the
metric in Gauss normal form and far away
from the source was simply the Neumann
boundary condition for the components of the metric.
We can derive the Israel matching condition for the
components of the metric in isotropic coordinates
by performing a coordinate transformation to the
Neumann boundary condition,
generated by $u$ and $v$, subjected to
constraints (\eqtevekidhs). Furthermore with the
identifications (\eqvekid) we derive the Israel
matching condition in the isotropic coordinates
$$
{{\partial\alpha}\over {\partial\eta}}|_{\eta \to 0}=
{{\partial\beta}\over
{\partial\eta}}|_{\eta\to 0}
=-{\kappa}f|_{\eta\to 0}
\numbereq\name{\eqsterios}
$$
It is straightforward then to verify that our linearized
solution in isotropic coordinates satisfies the Israel
matching condition. For example
$$
\eqalign{
{{\partial\alpha}\over {\partial\eta}}|_{\eta \to 0}=
k{\zeta}{{\partial\alpha}\over {\partial\zeta}}|_{\zeta
\to {1\over {\kappa}}}&=-2k{\zeta}
{{\partial\Phi}\over {\partial\zeta}}|_{\zeta
\to {1\over {\kappa}}}+k{\zeta}
{{\partial\Psi}\over {\partial\zeta}}|_{\zeta
\to {1\over {\kappa}}}=\cr
&=-{{4GM{\kappa}}\over {3{\pi}}}
\int_0^\infty dpp
{j_0(p{\rho})}{{{K_0(\hat p)}}\over {K_1(\hat p)}}
=-{\kappa}f(\rho, {1\over {\kappa}}).\cr}
\numbereq\name{\eqskenter}
$$
We have thus derived an exact solution to linear order in
the gravitational coupling which is both free from singularities
and satisfies the Israel matching condition.

\newsection Extensions of the Horizon in the Bulk.

Beyond the linear approximation, the Einstein equations
are difficult to solve. The presence of Bessel functions
in the linear approximation makes it implausible that a closed
form exact solution exists. Nevertheless, some rigorous statements
can still be made. An interesting issue is the 5-D extension
of the 4-D Schwarzschild horizon for $\kappa r>>1$.
A discussion on black holes in the brane world can
be found in \ref{\bla}{R. Emparan,
G. Horowitz and R. Myers, \jhep0001 (2000) 007;
T. Shiromizu and M. Shibata, \pr62 (2000) 127502;
A. Chamblin, H. Reall, H. Shinkai
and T. Shiromizu, \pr63 (2001) 064015;
N. Dadhich, R. Maartens, P. Papadopoulos
and V. Rezania, \pl487 (2000) 1.} In a previous
paper \ref{\ion}{I. Giannakis and H. C. Ren, {\it
Possible Extensions of the 4-D Schwarzschild Horizon
in the Brane World}, hep-th/0010183, to appear in Phys. Rev. D}
we showed that within the Gauss-normal form of the metric,
such an extension can only take a tubular shape.
We shall examine this problem for a general diagonal form
of the metric
$$
ds^2=-e^adt^2+e^bdr^2+e^cr^2d\Omega^2+e^fdy^2.
\numbereq\name{\eqeikosi}
$$
where we have absorbed the conformal factor $e^{-2{\kappa}y}$
into the definition of $a$, $b$ and $c$.
The non-zero components of the Ricci tensor are presented
in the Appendix. What we need here is
$$
\eqalign{
R_{yy}&={1\over 2}(\ddot a+\ddot b+2\ddot c)
+{1\over 4}(\dot a^2
+\dot b^2+2\dot c^2)\cr
&+{1\over 2}e^{f-b}[f^{\prime\prime}
+{1\over 2}({4\over r}+a^{\prime}-b^{\prime}+2c^{\prime}
+f^{\prime})f^{\prime}]
-{1\over 4}(\dot a+\dot b+2\dot c){\dot f}\cr}
\numbereq\name{\eqdekaenea}
$$
Let $H(r, y)=0$ describe the trajectory of the horizon on the $r-y$
parametric plane. Consider a point
$P(r_0, y_0)$ on the trajectory at which the unit normal to the
horizon $\vec n=(\cos\alpha, \sin\alpha)$ with
$$
\cos\alpha = {1\over {\Delta}}\Big({{\partial H}\over {\partial r}}\Big)_P,
\qquad \sin\alpha ={1\over {\Delta}}
\Big({{\partial H}\over {\partial y}}\Big)_P
\numbereq\name{\eqthanou}
$$
and 
$$
\Delta = \sqrt{\Big({{\partial H}\over {\partial r}}\Big)_P^2
+\Big({{\partial H}\over {\partial y}}\Big)_P^2}.
\numbereq\name{\eqxantop}
$$
In the neighbourhood of $P$ we may introduce the normal and
tangent  coordinates $\xi$  and $\eta$
$$
\eqalign{
\xi&=(r-r_0)\cos\alpha+(y-y_0)\sin\alpha\cr
\eta&=-(r-r_0)\sin\alpha+(y-y_0)\cos\alpha.\cr}
\numbereq\name{\eqkarnat}
$$
For the standard form of the Schwarzschild horizon, we expect
that
$$
a\simeq \ln\xi, \qquad b\simeq \ln\xi
\numbereq\name{\eqverouli}
$$
and $c$ nonsingular as $\xi, \eta \to 0$. Assuming
that $ f\simeq n\ln\xi$ with $n$ an even integer we have
$$
\eqalign{
\dot a&\simeq {1\over {\xi}}\sin\alpha, \qquad \dot b\simeq
-{1\over{\xi}}\sin\alpha, \qquad \dot f\simeq{n\over\xi}\sin\alpha\cr
a^\prime&\simeq {1\over {\xi}}\cos\alpha, \qquad b^\prime\simeq
-{1\over{\xi}}\cos\alpha, \qquad f^\prime\simeq{n\over\xi}\cos\alpha\cr}
\numbereq\name{\eqpyrros}
$$
and
$$
\eqalign{
{\ddot a}&\simeq -{1\over {\xi}^2}\sin^2\alpha, \qquad
{\ddot b}\simeq
{1\over{\xi}^2}\sin^2\alpha,
\qquad {\ddot f}\simeq-{n\over{\xi}^2}\sin^2\alpha \cr
a^{\prime\prime}&\simeq -{1\over {\xi}^2}\cos^2\alpha,
\qquad b^{\prime\prime}\simeq
{1\over{\xi}^2}\cos^2\alpha, \qquad f^{\prime\prime}
\simeq-{n\over{\xi}^2}\cos^2\alpha\cr}
\numbereq\name{\eqpyrrote}
$$
The Einstein equation $R_{yy}-4{\kappa}^2e^f=0$ demands
the absence of the singularity on the left hand side.
For $n<-1$, the leading singularity stems from the term in the
bracket of equation (\eqdekaenea), i.e.,
$$
R_{yy}\sim n^2{\xi}^{-1+n}\cos^2\alpha
\numbereq\name{\eqdekaoktw}
$$
which implies $\cos\alpha=0$. This horizon is parallel to the
brane and will not join the 4D Schwarzschild horizon on the brane.
For $n>-1$, the leading singularity comes from the second term
of equation (\eqdekaenea), i.e.,
$$
R_{yy}\sim {1\over {2{\xi}^2}}\sin^2\alpha
\numbereq\name{\eqdekaefta}
$$
and we have $\sin\alpha=0$ which implies a tubular horizon.
Therefore we conclude that the 5-D extension of the standard
Schwarzschild horizon with any diagonal form of the metric
is tubular.

The situation changes, however, for the 5-D extension of the
isotropic form of the 4-D Schwarzschild horizon, for which
$a\simeq 2\ln\xi$ and $b=c$ nonsingular. With the assumption
that $f=n\ln\xi$, we find
$$
R_{yy} \sim Cn^2{\xi}^{n-2}\cos^2\alpha+Dn{\xi}^{-2}\sin^2\alpha
\numbereq\name{\eqdekaexi}
$$
as $\xi, \eta \to 0$ and $C$, $D$ being constants.
It follows that $\cos\alpha=0$ for $n=0$ and $\sin\alpha=0$
for $n>0$. There is no restriction on the horizon shape for
$n=0$ and in this case $f$ is nonsingular as well. The coexistence
of a straight brane and the $AdS_5$ horizon within the coordinate
patch specified by $b=c$, as it is revealed by the linear gravity
analysis, implies a non-trivial shape- including a pancake) for
the 5-D extension of the 4-D horizon.

\newsection Concluding Remarks.

In this section we will recapitulate what we have done in this paper. We
considered a static, axially symmetric
metric in $D=4+1$ dimensions in the Randall-Sundrum framework
in isotropic coordinates and derived the
Einstein equations. We subsequently found an exact solution
of the equations
to first order in the gravitational coupling.
This solution
is free from singularities as $y \to \infty$ and
satisfies the Israel matching condition at $y=0$. It describes
the gravitational field of a spherically
symmetric mass distribution confined on
the {\it physical brane} which is straight in the isotropic coordinates.
At distances far away from the material source and
on the {\it physical brane} this solution coincides
with the four dimensional
Schwarzschild metric in isotropic coordinates. Thus
we confirmed that all tests of Linearized General Relativity are
satisfied. We have also found that the vacuum
Einstein equations together with 
the form of the metric place fairly stringent restrictions on 
the shape of the event horizon. More specifically we showed that
the $5-D$ extension of the standard Schwarzschild horizon with
any diagonal form of the metric is tubular while there is
no restriction for the $5-D$
extension of the Schwarzschild horizon in isotropic coordinates.

Finally we would like to speculate on the structure of the solution
in isotropic coordinates beyond the linearized approximation.
In accordance with the discussion in the previous section, the
metric near the horizon can be approximated by
$$
ds^2=-{\xi}^2dt^2+d{\xi}^2+d{\vec{\zeta}}^2
\numbereq\name{\eqmakiko}
$$
where $d{\vec{\zeta}}^2$ is a $3-D$ Euclidean metric. We notice
that the determinant of the metric vanishes on the
horizon. The transformation to a local inertial frame,
$T={\xi}{\cosh t}$ and $X={\xi}{\sinh t}$ does not
cover the region with $T^2>X^2$ and therefore the
isotropic coordinate patch does not cover the entire spacetime.

To be more specific we consider the unrealistic case
$\kappa GM<<1$. The $4-D$ brane gravity joins the $5-D$
gravity in the region where the linearized approximation
is still valid. For $\rho$, $\eta$ satisfying the inequality
$$
({GM\over {\kappa}})^{1\over 2}<< l << ({GM\over {\kappa}^2})^{1\over 3}
\numbereq\name{\equchida}
$$
with $l=({\rho}^2+{\eta}^2)^{1\over 2}$, the metric (\eqexi),
with $\alpha$, $\beta$ and $f$ given by (\eqartis), becomes
approximately
$$
ds^2=-(1-{8GM\over {3{\pi}{\kappa}l^2}})dt^2+
(1+{{4GM}\over {3{\pi}{\kappa}l^2}})(d{\rho}^2+{\rho}^2d\Omega^2+d{\eta}^2).
\numbereq\name{\eqtudor}
$$
The lower bound of (\equchida) results from the linear approximation
while the upper one from ignoring the conformal factor
$e^{-2{\kappa}{\eta}}$. As we go beyond the lower bound of (\equchida),
$\alpha$, $\beta$ and $f$ are expected to coincide with the
corresponding expressions of the isotropic form
of the $5-D$ Schwarzschild metric
$$
ds^2=-({{l^2-l_{0}^2}\over {{l^2+l_{0}^2}}})^2dt^2+
(1+{l_{0}^2\over l^2})^2(d{\rho}^2+{\rho}^2d\Omega^2+d{\eta}^2).
\numbereq\name{\eqtudoris}
$$
with $l_{0}^2={2GM\over {3{\pi}{\kappa}}}$. In this case
we have a trivial case of a closed $5-D$ horizon. It is well
known for $\kappa=0$ that the isotropic coordinates for $l<l_0$
represent a copy of the spacetime with $l>l_0$ as it is obvious
from the mapping $l \to l'={l_{0}^2\over l}$ which amounts to
$$
\rho \to \rho'={{l_{0}^2{\rho}}\over {{\rho}^2+{\eta}^2}},
\qquad \eta \to \eta'={{l_{0}^2{\eta}}\over {{\rho}^2+{\eta}^2}}.
\numbereq\name{\eqarvion}
$$
The approximation of ignoring $\kappa$ continues to hold with
decreasing $l$ for $l<l_0$ until $\alpha \sim 2{\kappa}{\eta}$
beyond which we shall find ourselves in the region where
$4-D$ gravity dominates again. Therefore the physical black hole solution
specified by the isotropic coordinates for ${\kappa}GM << 1$
describes only the spacetime outside the Schwarzschild horizon,
as for the case $\kappa=0$. This property might pertain as the
parameter ${\kappa}GM$ is continuated to more realistic regions,
i. e. ${\kappa}GM >> 1$. In that case, the solution to the brane
world Einstein equations expressed in isotropic coordinates
never probes the spacetime interior to the Schwarzschild
horizon of a physical black hole, in particular the
coordinate patch never extends to the curvature singularity.

\vskip .1in
\noindent
{\bf Acknowledgments.} \vskip .01in \noindent
We are grateful to Z. Kakushadze for communicating with
us his work on gauge fixing [\kaku].
We would like also to thank N. Khuri, J. T. Liu, B. Morariu
and A. Polychronakos,
for useful discussions.
This work was supported in part by the Department of Energy under Contract
Number DE-FG02-91ER40651-TASK B.

\newsection Appendix.

In this appendix, we tabulate all non-zero components of
the Ricci tensor for the general diagonal
metric {\it ans\"atz\/} (\eqeikosi).
Our sign conventions follows that of reference [\wei]. We find
$$
\eqalign{
R_{tt}&={1\over 2}e^{a-b}\Big[-a^{\prime\prime}
-{2\over r}a^\prime
+{1\over 2}a^\prime(-a^\prime+b^\prime
-2c^\prime)-{1\over 2}a^{\prime}f^{\prime}\Big]+{1\over 2}
e^{a-f}\Big[-\ddot a-{1\over 2}\dot a(\dot a+\dot b+2\dot c\cr
&-\dot f)\Big]\cr
R_{rr}&={1\over 2}a^{\prime\prime}+c^{\prime\prime}
+{1\over 2}f^{\prime\prime}
-{1\over r}b^\prime+{2\over r}c^\prime
+{1\over 4}a^\prime(a^\prime-b^\prime)-{1\over 2}
c^\prime(b^\prime-c^\prime)+{1\over 4}f^{\prime 2}-{1\over 4}b^{\prime}
f^{\prime}\cr
&+{1\over 2}e^{b-f}\Big[\ddot b+
{1\over 2}\dot b(\dot a+\dot b+2\dot c-\dot f\Big]\cr
R_{\theta\theta}&=-1+e^{c-b}+r^2e^{c-b}\Big[{1\over 2}
c^{\prime\prime}+{2\over r}c^\prime+{{a^\prime-b^\prime+f^\prime}\over 2r}
+{1\over 2}c^{\prime2}
+{1\over 4}(a^\prime-b^\prime)c^\prime+{1\over 4}c^{\prime}f^{\prime}
\Big]\cr
&+e^{c-f}r^2\Big[{1\over 2}
\ddot c+{1\over 4}\dot c(\dot a+\dot b
+2\dot c-\dot f)\Big]\cr
R_{\phi\phi}&=R_{\theta\theta}{\sin^2{\theta}}\cr
R_{yy}&={1\over 2}(
\ddot a+\ddot b+2\ddot c)+{1\over 4}(\dot a^2
+\dot b^2+2\dot c^2)-{1\over 4}{\dot f}(\dot a+\dot b+2\dot c)\cr
&+{1\over 2}e^{f-b}\Big[f^{\prime\prime}+{1\over 2}
({4\over r}+a^{\prime}-b^{\prime}+2c^{\prime}+f^{\prime})f^{\prime}\Big]\cr
R_{ry}&=R_{yr}={1\over 2}\Big[
\dot a^\prime+2\dot c^\prime-{2\over r}(\dot b-\dot c)+{1\over 2}a^\prime
(\dot a-\dot b)-c^\prime(\dot b-\dot c)
-{1\over 2}(\dot a+2\dot c)f^{\prime}\Big]\cr}
\numbereq\name{\eqakrathtos}
$$
The Gauss normal {\it ans\"atz\/} corresponds to the identification
$$
a \to a-2{\kappa}y, \qquad b \to b-2{\kappa}y, \qquad
c \to c-2{\kappa}y, \qquad f=0,
\numbereq\name{\eqaseri}
$$
while the isotropic form of the metric, employed in this
work corresponds to $r \to \rho$, $y \to \eta$ and
$$
a={\alpha}-2{\kappa}{\eta}, \qquad b={\beta}
-2{\kappa}{\eta}, \qquad
c=-2{\kappa}{\eta}, \qquad f \ne 0.
\numbereq\name{\eqaserko}
$$

\noindent
\immediate\closeout1
\bigbreak\bigskip

\line{\twelvebf References. \hfil}
\nobreak\medskip\vskip\parskip

\input refs
\vfil\end

\bye